\documentclass[12pt,twoside]{article}
\usepackage{amssymb,amsmath,mathrsfs,hogg_endnotes,natbib}
\usepackage{float,graphicx}

\newcommand{\notenglish}[1]{\textsl{#1}}

\newcommand{\etal}{\notenglish{et al.}}

\newcommand{\documentname}{document}
\newcommand{\documentnames}{\documentname s}
\newcommand{\sectionname}{Section}
\newcommand{\Equationname}{Equation}
\newcommand{\equationname}{equation}
\newcommand{\equationnames}{\equationname s}

\newcommand{\problemname}{Exercise}

\newcommand{\notename}{note}

\newcommand{\note}[1]{\endnote{#1}}
\def\enotesize{\normalsize}

\newcounter{problem}
\newenvironment{problem}{\paragraph{\problemname~\theproblem:}\refstepcounter{problem}}{}
\newcommand{\affil}[1]{{\footnotesize\textsl{#1}}}
\renewcommand{\and}{{\footnotesize{and}}}

\newcommand{\setofall}[3]{\{{#1}\}_{{#2}}^{{#3}}}

\newcommand{\dd}{\mathrm{d}}
\newcommand{\given}{\,|\,}

\renewcommand{\MakeUppercase}[1]{#1}
\begin{document}
\thispagestyle{plain}\raggedbottom
\section*{Data analysis recipes:\\
  Probability calculus for inference\footnotemark}

\footnotetext{%
  The \notename s begin on page~\pageref{note:first}, including the
  license\note{\label{note:first}%
    Copyright 2012 David W. Hogg (NYU).  You may copy and distribute this
    document provided that you make no changes to it whatsoever.}
  and the acknowledgements\note{%
    It is a pleasure to thank
      Jo Bovy (IAS),
      Kyle Cranmer (NYU),
      Phil Marshall (Oxford),
      Hans-Walter Rix (MPIA),
      and Dan Weisz (UW)
    for discussions and comments.  This
    research was partially supported by the US National Aeronautics
    and Space Administration and National Science Foundation.}}

\noindent
David~W.~Hogg\\
\affil{Center~for~Cosmology~and~Particle~Physics, Department~of~Physics, New York University}\\
\and~\affil{Max-Planck-Institut f\"ur Astronomie, Heidelberg}

\begin{abstract}
In this pedagogical text aimed at those wanting to start thinking
about or brush up on probabilistic inference, I review the rules by
which probability distribution functions can (and cannot) be
combined. I connect these rules to the operations performed in
probabilistic data analysis.  Dimensional analysis is emphasized as a
valuable tool for helping to construct non-wrong probabilistic
statements.  The applications of probability calculus in constructing
likelihoods, marginalized likelihoods, posterior probabilities, and
posterior predictions are all discussed.
\end{abstract}

\noindent
When constructing the plan or basis for a probabilistic inference---a
data analysis making use of likelihoods and also prior probabilities
and posterior probabilities and marginalization of nuisance
parameters---the options are \emph{incredibly strongly constrained} by
the simple rules of probability calculus.  That is, there are only
certain ways that probability distribution functions (``pdfs'' in what
follows) \emph{can} be combined to make new pdfs, compute expectation
values, or make other kinds of non-wrong statements.  For this reason,
it behooves the data analyst to have good familiarity and facility
with these rules.

Formally, probability calculus is extremely simple.  However, it is
not always part of a physicist's (or biologist's or chemist's or
economist's) education.  That motivates this \documentname.

For space and specificity---and because it is useful for most
problems I encounter---I will focus on continuous variables (think
``model parameters'' for the purposes of inference) rather than
binary, integer, or discrete parameters, although I might say one or
two things here or there.  I also won't always be specific about the
domain of variables (for example whether a variable $a$ is defined on
$0<a<1$ or $0<a<\infty$ or $-\infty<a<\infty$); the limits of
integrals---all of which are definite---will usually be
implicit.\note{In my world, \emph{all integrals are definite
    integrals}.  The integral is never just the anti-derivative, it is
  always a definite, finite, area or volume or measure.  That has to
  do, somehow, with the fact that the rules of probability calculus
  are an application of the rules of measure theory.}  Along the way,
a key idea I want to convey is---and this shows that I come from
physics---that it is very helpful to think about the units or
dimensions of probability quantities.

I learned the material here by using it; probability calculus is so
simple and constraining, you can truly learn it as you go.  However if
you want more information or discussion about any of the subjects
below, \cite{sivia} provide a book-length, practical introduction and
the first few chapters of \cite{jaynes} make a great and exceedingly
idiosyncratic read.

\section{Generalities}

A probability distribution function has units or dimensions.  Don't
ignore them.  For example, if you have a continuous parameter $a$, and
a pdf $p(a)$ for $a$, it must obey the normalization condition
\begin{eqnarray}\displaystyle
1 &=& \int p(a)\,\dd a
\quad ,
\end{eqnarray}
where the limits of the integral should be thought of as going over
the entire domain of $a$.  This (along with, perhaps, $p(a)\geq 0$
everywhere) is almost the \emph{definition} of a pdf, from my
(pragmatic, informal) point of view.  This normalization condition
shows that $p(a)$ has units of $a^{-1}$.  Nothing else would integrate
properly to a dimensionless result.  Even if $a$ is a
multi-dimensional vector or list or tensor or field or even point in
function space, the pdf must have units of $a^{-1}$.

In the multi-dimensional case, the units of $a^{-1}$ are found by
taking the product of all the units of all the dimensions.  So, for
example, if $a$ is a six-dimensional phase-space position in
three-dimensional space (three cartesian position components measured
in m and three cartesian momentum components measured in
kg\,m\,s$^{-1}$), the units of $p(a)$ would be
kg$^{-3}$\,m$^{-6}$\,s$^3$.

Most problems we will encounter will have multiple parameters; even if
we \emph{condition} $p(a)$ on some particular value of another
parameter $b$, that is, ask for the pdf for $a$ \emph{given} that $b$
has a particular, known value to make $p(a \given b)$ (read ``the pdf
for $a$ given $b$''), it must obey the same normalization
\begin{eqnarray}\displaystyle
1 &=& \int p(a \given b)\,\dd a
\quad ,
\end{eqnarray}
but you can \emph{absolutely never do} the integral
\begin{eqnarray}\displaystyle\label{eq:wrong1}
\mbox{\textbf{wrong:}} & & \int p(a \given b)\,\dd b
\end{eqnarray}
because that integral would have units of $a^{-1}\,b$, which is (for
our purposes) absurd.\note{To see that the units in
  \equationname~(\ref{eq:wrong1}) are $a^{-1}\,b$, you have to see
  that the \emph{integration operator} $\int\dd b$ itself has units of
  $b$ (just as the derivative operator $\dd/\dd b$ has units of
  $b^{-1}$).  Why are the units $a^{-1}\,b$ absurd?  Well, they might
  not be absurd units in general, but they aren't what we want when
  we integrate a probability distribution.}

If you have a probability distribution for two things (``the pdf for
$a$ and $b$''), you can always factorize it into two distributions,
one for $a$, and one for $b$ given $a$ or the other way around:
\begin{eqnarray}\displaystyle
p(a, b) &=& p(a)\,p(b \given a)
\\
p(a, b) &=& p(a \given b)\,p(b)
\quad ,
\end{eqnarray}
where the units of both sides of both equations are $a^{-1}\,b^{-1}$.
These two factorizations taken together lead to what is sometimes
called ``Bayes's theorem'', or
\begin{eqnarray}\displaystyle
p(a \given b) &=& \frac{p(b \given a)\,p(a)}{p(b)}
\quad ,
\end{eqnarray}
where the units are just $a^{-1}$ (the $b^{-1}$ units cancel out top
and bottom), and the ``divide by $p(b)$'' aspect of that gives many
philosophers and mathematicians the chills (though certainly not
me).\note{Division by zero is a huge danger---in principle---when
  applying Bayes's theorem.  In practice, if there is support for the
  model in your data, or support for the data in your model, you don't
  hit any zeros.  This sounds a little crazy, but if you have data
  with unmodeled outliers---for example if your noise model can't
  handle enormous data excursions caused by rare events---you can
  easily get real data sets that have vanishing probability in a
  na\"ive model.}  Conditional probabilities factor just the same as
unconditional ones (and many will tell you that there is no such thing
as an unconditional probability\note{There are no unconditional
  probabilities!  This is because whenever \emph{in practice} you
  calculate a probability or a pdf, you are always making strong
  assumptions.  Your probabilities are all conditioned on these
  assumptions.}); they factor like this:
\begin{eqnarray}\displaystyle
p(a, b \given c) &=& p(a \given c)\,p(b \given a, c)
\\
p(a, b \given c) &=& p(a \given b, c)\,p(b \given c)
\\
p(a \given b, c) &=& \frac{p(b \given a, c)\,p(a \given c)}{p(b \given c)}
\quad ,
\end{eqnarray}
where the condition $c$ must be carried through all the terms; the
whole right-hand side must be conditioned on $c$ if the left-hand side
is.  Again, there was Bayes's theorem, and you can see its role in
conversions of one kind of conditional probability into another.  For
technical reasons,\note{The ``technical reason'' here that I treat the
  denominator in \equationname~(\ref{eq:bayes}) as a renormalization constant $Z$ is that when we
  perform Markov-Chain Monte Carlo sampling methods to obtain samples
  from $p(a \given b, c)$, we will not need to know $Z$ at all; it is
  usually (in practice) hard to compute and often unnecessary.} I
usually write Bayes's theorem like this:
\begin{eqnarray}\displaystyle\label{eq:bayes}
p(a \given b, c) &=& \frac{1}{Z}\,p(b \given a, c)\,p(a \given c)
\\
Z &\equiv& \int p(b \given a, c)\,p(a \given c)\,\dd a
\quad .
\end{eqnarray}

Here are things you \emph{can't} do:
\begin{eqnarray}\displaystyle
\mbox{\textbf{wrong:}} & & p(a \given b, c)\,p(b \given a, c)
\\
\mbox{\textbf{wrong:}} & & p(a \given b, c)\,p(a \given c)
\quad;
\end{eqnarray}
the first over-conditions (it is not a factorization of anything
possible) and the second has units of $a^{-2}$, which is absurd (for
our purposes).  Know these and \emph{don't do them}.

One important and confusing point about all this: The terminology used
throughout this \documentname\ \emph{enormously overloads} the symbol
$p(\cdot)$.  That is, we are using, in each line of this discussion,
the function $p(\cdot)$ to mean something different; it's meaning is
set by the letters used in its arguments.  That is a nomenclatural
abomination.\note{Serious, non-ambiguous mathematicians often
  distinguish between the name of the variable and the name of a draw
  from the probability distribution for the variable, and then instead
  of $p(a \given b, c)$ they can write things like $p_{A,B=b,C=c}(a)$,
  which are unambiguous (or far less ambiguous).  This permits, for
  example, another thing $q$ to be drawn from the same distribution as
  $a$ by a notation like $p_{A,B=b,C=c}(q)$.  In our (very bad)
  notation $p(q \given b, c)$ would in general be different from $p(a
  \given b, c)$ because we take the meaning of the function from the
  names of the argument variables.  See why that is very bad?  Some
  explicit subscripting policy seems like much better practice and we
  should probably all adopt something like it, though I won't
  here.\par Another good idea is to make the probability equations be
  about statements, so instead of writing $p(a \given b, c)$ you write
  $p(A=a \given B=b, C=c)$.  This is unambiguous---you can write
  useful terms like $p(A=q \given B=b, C=c)$ to deal with the $a$, $q$
  problem---but it makes the equations big and long.}  I apologize,
and encourage my readers to do things that aren't so ambiguous (like
maybe add informative subscripts), but it is so standard in our
business that I won't change (for now).

The theory of continuous pdfs is measure theory; measure theory (for
me, anyway\note{Have you noticed that I am \emph{not} a
  mathematician?}) is the theory of things in which you can do
integrals.  You can \emph{integrate out} or \emph{marginalize away}
variables you want to get rid of (or, in what follows, \emph{not}
infer) by integrals that look like
\begin{eqnarray}\displaystyle
p(a \given c) &=& \int p(a, b \given c)\,\dd b
\\\label{eq:marginalize}
p(a \given c) &=& \int p(a \given b, c)\,p(b \given c)\,\dd b
\quad ,
\end{eqnarray}
where the second is a factorized version of the first.  Once again the
integrals go over the entire domain of $b$ in each case, and again if
the left-hand side is conditioned on $c$, then everything on the
right-hand side must be also.  This equation is a natural consequence
of the things written above and dimensional analysis.  Recall that
because $b$ is some kind of arbitrary, possibly very high-dimensional
mathematical object, these integrals can be extremely daunting in
practice (see below).  Sometimes equations like (\ref{eq:marginalize})
can be written
\begin{eqnarray}\displaystyle
p(a \given c) &=& \int p(a \given b)\,p(b \given c)\,\dd b
\quad ,
\end{eqnarray}
where the dependence of $p(a \given b)$ on $c$ has been dropped.  This
is only permitted if it \emph{happens to be the case} that $p(a \given
b, c)$ doesn't, in practice, depend on $c$.  The dependence on $c$ is
really there (in some sense), it just might be trivial or null.

In rare cases, you can get factorizations that look like this:
\begin{eqnarray}\displaystyle
p(a, b \given c) &=& p(a \given c)\,p(b \given c)
\quad ;
\end{eqnarray}
this factorization doesn't have the pdf for $a$ depend on $b$ or vice
versa.  When this happens---and it is rare---it says that $a$ and $b$
are ``independent'' (at least conditional on $c$).\note{The word
  ``independent'' has many meanings in different contexts of
  mathematics and probability; I will avoid it in what follows, except
  in context of independently drawn data points in a generative model
  of data.  I prefer the word ``separable'' for this situation,
  because I think it is less ambiguous.}  In many cases of data
analysis, in which we want probabilistic models of data sets, we often
have some number $N$ of data $a_n$ (indexed by $n$) each of which is
independent in this sense.  If for each datum $a_n$ you can write down
a probability distribution $p(a_n \given c)$, then the probability of
the full data set---when the data are independent---is simply the
product of the individual data-point probabilities:
\begin{eqnarray}\displaystyle
p(\setofall{a_n}{n=1}{N} \given c) &=& \prod_{n=1}^N p(a_n \given c)
\quad .
\end{eqnarray}
This is the definition of ``independent data''.  If all the functions
$p(a_n \given c)$ are identical---that is, if the functions don't
depend on $n$---we say the data are ``iid'' or ``independent and
identically distributed''.  However, this will not be true in general
in data analysis.  Real data are heteroscedastic at the very
least.\note{I love the word ``heteroscedastic''.  It means ``having
  heterogeneous noise properties'', or that you can't treat every data
  point as being drawn from the same noise model.  Heteroscedasticity
  is a property of every data set I have ever used.}

I am writing here mainly about continuous variables, but one thing
that comes up frequently in data analysis is the idea of a ``mixture
model'' in which data are produced by two (or more) qualitatively
different processes (some data are good, and some are bad, for
example) that have different relative probabilities.  When a variable
($b$, say) is discrete, the marginalization integral corresponding to
\equationname~(\ref{eq:marginalize}) becomes a sum
\begin{eqnarray}\displaystyle
p(a \given c) &=& \sum_b p(a \given b, c)\,p(b \given c)
\quad ,
\end{eqnarray}
and the normalization of $p(b \given c)$ becomes
\begin{eqnarray}\displaystyle
1 &=& \sum_b p(b \given c)
\quad ;
\end{eqnarray}
in both sums, the sum is implicitly over the (countable number of)
possible states of $b$.\note{In many sources, when a variable $b$ is
  made discrete, because the pdf $p(b\given a)$ becomes just a set of
  probabilities, the symbol ``$p$'' is often changed to ``$P$''.
  That's not crazy; anything that makes this ambiguous nomenclature
  less ambiguous is good.}

If you have a conditional pdf for $a$, for example $p(a \given c)$,
and you want to know the expectation value $E(a \given c)$ of $a$
under this pdf (which would be, for example, something like the mean
value of $a$ you would get if you drew many draws from the conditional
pdf), you just integrate
\begin{eqnarray}\displaystyle
E(a \given c) &=& \int a\,p(a \given c)\,\dd a
\quad .
\end{eqnarray}
This generalizes to any function $f(a)$ of $a$:
\begin{eqnarray}\displaystyle
E(f \given c) &=& \int f(a)\,p(a \given c)\,\dd a
\quad .
\end{eqnarray}
You can see the marginalization integral (\ref{eq:marginalize}) that
converts $p(a \given b, c)$ into $p(a \given c)$ as providing the
\emph{expectation value} of $p(a \given b, c)$ under the conditional
pdf $p(b \given c)$.  That's deep and relevant for what follows.

\begin{problem}
You have conditional pdfs $p(a \given d)$, $p(b \given a, d)$, and
$p(c \given a, b, d)$.  Write expressions for $p(a, b \given d)$, $p(b
\given d)$, and $p(a \given c, d)$.
\end{problem}

\begin{problem}
You have conditional pdfs $p(a \given b, c)$ and $p(a \given c)$
expressed or computable for any values of $a$, $b$, and $c$.  You are
not permitted to multiply these together, of course.  But can you use
them to construct the conditional pdf $p(b \given a, c)$ or $p(b
\given c)$?  Did you have to make any assumptions?
\end{problem}

\begin{problem}
You have conditional pdfs $p(a \given c)$ and $p(b \given c)$
expressed or computable for any values of $a$, $b$, and $c$.  Can you
use them to construct the conditional pdf $p(a \given b, c)$?
\end{problem}

\begin{problem}
You have a function $g(b)$ that is a function only of $b$.  You have
conditional pdfs $p(a \given c)$ and $p(b \given a, c)$.  What is the
expectation value $E(g \given c)$ for $g$ conditional on $c$ but
\emph{not} conditional on $a$?
\end{problem}

\begin{problem}
Take the integral on the right-hand side of
\equationname~(\ref{eq:marginalize}) and replace the ``$\dd b$'' with
a ``$\dd a$''.  Is it permissible to do this integral?  Why or why
not?  If it \emph{is} permissible, what do you get?
\end{problem}

\section{Likelihoods}

Imagine you have $N$ data points or measurements $D_n$ of some kind,
possibly times or temperatures or brightnesses.  I will say that you
have a ``generative model'' of data point $n$ if you can write down or
calculate a pdf $p(D_n \given \theta, I)$ for the measurement $D_n$,
conditional on a vector or list $\theta$ of parameters and a (possibly
large) number of other things $I$ (``prior information'') on which the
$D_n$ pdf depends, such as assumptions, or approximations, or
knowledge about the noise process, or so on.  If all the data points
are independently drawn (that would be one of the assumptions in $I$),
then the pdf for the full data set $\setofall{D_n}{n=1}{N}$ is just
the product
\begin{eqnarray}\displaystyle\label{eq:likelihood}
p(\setofall{D_n}{n=1}{N} \given \theta, I) &=& \prod_{n=1}^N p(D_n \given \theta, I)
\quad .
\end{eqnarray}
(This requires the data to be independent, but not necessarily iid.)
When the pdf of the data is thought of as being a \emph{function} of
the parameters at \emph{fixed} data, the pdf of the data given the
parameters is called the ``likelihood'' (for historical reasons I
don't care about).  In general, in contexts in which the data are
thought of as being fixed and the parameters are thought of as
variable, any kind of conditional pdf for the data---conditional on
parameters---is called a likelihood ``for the parameters'' even though
it is a pdf ``for the data''.\note{I hate the terminology
  ``likelihood'' and ``likelihood for parameters'' but again, it is so
  entrenched, it would be a disservice to the reader not to use it.
  If you really want to go down the rabbit hole, \citet{jaynes} calls
  the pdf for the data a ``frequency distribution'' not a
  ``probability distribution'' because he \emph{always} sees the data
  as being \emph{fixed} and reserves the word ``probability'' for the
  things that are unknown (that is, being inferred).}

Now imagine that the parameters divide into two groups.  One group
$\theta$ are parameters of great interest, and another group $\alpha$
are of no interest.  The $\alpha$ parameters are nuisance parameters.
In this situation, the likelihood can be written
\begin{eqnarray}\displaystyle
p(\setofall{D_n}{n=1}{N} \given \theta, \alpha, I) &=& \prod_{n=1}^N p(D_n \given \theta, \alpha, I)
\quad .
\end{eqnarray}
If you want to make likelihood statements about the important
parameters $\theta$ without committing to anything regarding the
nuisance parameters $\alpha$, you can marginalize rather than infer
them.  You might be tempted to do
\begin{eqnarray}\displaystyle\label{eq:wrongmlikelihood}
\mbox{\textbf{wrong:}} & & \int p(\setofall{D_n}{n=1}{N} \given \theta, \alpha, I)\,\dd\alpha
\quad ,
\end{eqnarray}
but that's not allowed for dimensional arguments given in the previous
\sectionname.\note{The pseudo-marginalization operation in wrong
  \equationname~(\ref{eq:wrongmlikelihood}) has occasionally been done
  in the literature, and not led to disastrous results.  Why not?  It
  is because that operation, though not permitted, is very very
  similar to the operation you \emph{in fact} perform when you do the
  correct marginalization integration given in
  \equationname~(\ref{eq:mlikelihood}) but with a prior pdf $p(\alpha
  \given \theta, I)$ that is flat in $\alpha$.  Implicitly, \emph{any}
  marginalization or marginalization-like integral necessarily
  involves the choice of a prior pdf.}  In order to integrate over the
nuisances $\alpha$, something with units of $\alpha^{-1}$ needs to be
multiplied in---a pdf for the $\alpha$ of course:
\begin{eqnarray}\displaystyle\label{eq:mlikelihood}
p(\setofall{D_n}{n=1}{N} \given \theta, I) &=& \int p(\setofall{D_n}{n=1}{N} \given \theta, \alpha, I)\,p(\alpha \given \theta, I)\,\dd\alpha
\quad ,
\end{eqnarray}
where $p(\alpha \given \theta, I)$ is called the ``prior pdf'' for the
$\alpha$ and it \emph{can} depend (but doesn't \emph{need to} depend)
on the other parameters $\theta$ and the more general prior
information $I$.  This marginalization is incredibly useful but note
that it comes at a substantial cost: It required specifying a prior
pdf over parameters that, by assertion, you don't care about!\note{I
  will say a lot more about this in some subsequent \documentname\ in
  this series; in general in real-world problems you need to put a lot
  of care and attention into the parts of the problem that you
  \emph{don't} care about; it is a consequence of the need to make
  precise measurements in the parts of the problem that you \emph{do}
  care about.}  \Equationname~(\ref{eq:mlikelihood}) could be called a
``partially marginalized likelihood'' because it is a likelihood (a
pdf for the data) but it is conditional on fewer parameters than the
original, rawest, likelihood.

Sometimes I have heard concern that when you perform the
marginalization (\ref{eq:mlikelihood}), you are allowing the nuisance
parameters to ``have any values they like, whatsoever'' as if you are
somehow not constraining them.  It is true that you are not
\emph{inferring} the nuisance parameters, but you certainly are using
the data to limit their range, in the sense that the integral in
(\ref{eq:mlikelihood}) only gets significant weight where the
likelihood is large.  That is, if the data strongly constrain the
$\alpha$ to a narrow range of good values, then only those good values
are making any (significant) contribution to the marginalization
integral.  That's important!

Because the data were independent by assumption (the full-data
likelihood is a product of individual-datum likelihoods), you might be
tempted to do things like
\begin{eqnarray}\displaystyle\label{eq:wrong2}
\mbox{\textbf{wrong:}} & & \prod_{n=1}^N \left[\int p(D_n \given \theta, \alpha, I)\,p(\alpha \given \theta, I)\,\dd\alpha\right]
\quad .
\end{eqnarray}
This is wrong because if you do the integral \emph{inside} the
product, you end up doing the integral $N$ times over.  Or another way
to put it, although you don't want to infer the $\alpha$ parameters,
you want the support in the marginalization integral to be
consistently set by \emph{all} the data taken together, not
inconsistently set by each individual datum separately.

One thing that is often done with likelihoods (and one thing that many
audience members think, instinctively, when you mention the word
``likelihood'') is ``maximum likelihood''.  If all you want to write
down is your likelihood, and you want the ``best'' parameters $\theta$
given your data, you can find the parameters that maximize the
full-data likelihood.  The only really probabilistically responsible
use of the maximum-likelihood parameter value is---when coupled with a
likelihood width estimate---to make an approximate description of a
likelihood function.\note{A probabilistic reasoner returns
  probabilistic information, not just some ``best'' value.  That
  applies to frequentists and Bayesians alike.}  Of course a
maximum-likelihood value could in principle be useful on its own when
the data are \emph{so decisive} that there is (for the investigator's
purposes) no significant remaining uncertainty in the parameters.  I
have never seen that happen.  The key idea is that it is the
\emph{likelihood function} that is useful for inference, not some
parameter value suggested by that function.

To make all of the above concrete, we can consider a simple example,
where measurements $D_n$ are made at a set of ``horizontal'' postions
$x_n$.  Each measurement is of a ``vertical'' position $y_n$ but the
measurement is noisy, so the generative model looks like:
\begin{eqnarray}\displaystyle\label{eq:example0}
y_n &=& a\,x_n + b
\\
D_n &=& y_n + e_n
\\
p(e_n) &=& N(e_n \given 0, \sigma_n^2)
\\
p(D_n \given \theta, I) &=& N(D_n \given a\,x_n + b, \sigma_n^2)
\\\label{eq:examplelike}
p(\setofall{D_n}{n=1}{N} \given \theta, I) &=& \prod_{n=1}^N p(D_n \given \theta, I)
\\
\theta &\equiv& [a, b]
\\\label{eq:example1}
I &\equiv& [\setofall{x_n, \sigma_n^2}{n=1}{N}, \mbox{and so on\ldots} ]
\quad,
\end{eqnarray}
where the $y_n$ are the ``true values'' for the heights, which lie on
a straight line of slope $a$ and intercept $b$, the $e_n$ are noise
contributions, which are drawn independently from Gaussians $N(\cdot\given\cdot)$ with
zero means and variances $\sigma_n^2$, the likelihood is just a Gaussian
for each data point, there are two parameters, and the $x_n$ and
$\sigma_n^2$ values are considered prior information.\note{In general,
  the investigator has a lot of freedom in deciding what to treat as
  given, prior information, and what to treat as free parameters of
  the model.  If you don't trust your uncertainty variances
  $\sigma_n^2$, make them model parameters!  Same with the horizontal
  positions $x_n$.}

\begin{problem}
Show that the likelihood for the model given in
\equationnames~(\ref{eq:example0}) through (\ref{eq:example1}) can be
written in the form $Q\,\exp(-\chi^2/2)$, where $\chi^2$ is the
standard statistic for weighted least-squares problems.  On what does
$Q$ depend, and what are its dimensions?
\end{problem}

\begin{problem}
The likelihood in \equationname~(\ref{eq:examplelike}) is a product of
Gaussians in $D_n$.  At fixed data and $b$, what shape will it have in
the $a$ direction?  That is, what functional form will it have when
thought of as being a function of $a$?  You will have to use the
properties of Gaussians (and products of Gaussians).
\end{problem}

\section{Posterior probabilities}

A large fraction of the inference that is done in the quantitative
sciences can be left in the form of likelihoods and marginalized
likelihoods, and probably should be.\note{I will say more about what
  you both gain and lose by going beyond the likelihood in a
  subsequent \documentname\ in this series.  It relates to the battles
  between frequentists and Bayesians, which tend towards the boring
  and unproductive.}  However, there are many scientific questions the
answers to which require going beyond the likelihood---which is a pdf
for the data, conditional on the parameters---to a pdf for the
parameters, conditional on the data.

To illustrate, imagine that you want to make a probabilistic
prediction, given your data analysis. For one example, you might want
to predict what $\theta$ values you will find in a subsequent
experiment.  Or, for another, say some quantity $t$ of great interest
(like, say, the age of the Universe) is a function $t(\theta)$ of the
parameters and you want to predict the outcome you expect in some
future independent measurement of that same parameter (some other
experiment that measures---differently---the age of the Universe).
The distributions or expectation values for these predictions,
conditioned on your data, will require a pdf for the parameters or
functions of the parameters; that is, if you want the expectation $E(t
\given D)$, where for notational convenience I have defined
\begin{eqnarray}\displaystyle
D &\equiv& \setofall{D_n}{n=1}{N}
\quad ,
\end{eqnarray}
you need to do the integral
\begin{eqnarray}\displaystyle\label{eq:posteriorexpectation}
E(t \given D) &=& \int t(\theta)\,p(\theta \given D, I)\,\dd\theta
\quad ,
\end{eqnarray}
this in turn requires the pdf $p(\theta \given D, I)$ for the
parameters $\theta$ given the data.  This is called the ``posterior
pdf'' because it is the pdf you get \emph{after} digesting the data.

The posterior pdf is obtained by Bayes rule (\ref{eq:bayes})
\begin{eqnarray}\displaystyle\label{eq:posterior}
p(\theta \given D, I) &=& \frac{1}{Z}\,p(D \given \theta, I)\,p(\theta \given I)
\\
Z &\equiv& \int p(D \given \theta, I)\,p(\theta \given I)\,\dd\theta
\quad ,
\end{eqnarray}
where we had to introduce the prior pdf $p(\theta \given I)$ for the
parameters.  The prior can be thought of as the pdf for the parameters
\emph{before} you took the data $D$.  The prior pdf brings in new
assumptions but also new capabilities, because posterior expectation
values as in (\ref{eq:posteriorexpectation}) and other kinds of
probabilistic predictions become possible with its use.

Some notes about all this: \textsl{(a)}~Computation of expectation
values is not the only---or even the primary---use of posterior pdfs;
I was just using the expectation value as an example of \emph{why} you
might want the posterior.  You can use posterior pdfs to make
predictions that are themselves pdfs; this is in general the only way
to propagate the full posterior uncertainty remaining after your
experiment. \textsl{(b)}~The normalization $Z$ in
\equationname~(\ref{eq:posterior}) is a marginalization of a
likelihood; in fact it is a fully marginalized likelihood, and could
be written $p(D \given I)$.  It has many uses in model evaluation and
model averaging, to be discussed in subsequent \documentnames; it is
sometimes called ``the evidence'' though that is a very unspecific
term I don't like so much.  \textsl{(c)}~You can think of the
posterior expression (\ref{eq:posterior}) as being a ``belief
updating'' operation, in which you start with the prior pdf, multiply
in the likelihood (which probably makes it narrower, at least if your
data are useful) and re-normalize to make a posterior pdf.  There is a
``subjective'' attitude to take towards all this that makes the prior
and the posterior pdfs specific to the individual inferrer, while the
likelihood is (at least slightly) more objective.\note{Nothing in
  inference is ever truly \emph{objective} since every generative
  model involves making choices and assumptions and approximations.
  In particular, it involves choosing a (necessarily very limited)
  model space for inference and testing.  This informs my view that it
  is inconsistent---as a probabilistic data analyzer---to be a
  \emph{realist} \citep{hogg}.  All that said, the likelihood function
  is necessarily \emph{more} objective than the posterior pdf.}
\textsl{(d)}~Many committed Bayesians take the view that you
\emph{always} want a posterior pdf---that is, you are never satisfied
with a likelihood---and that you therefore \emph{must} always have a
prior, even if you don't think you do.  That view is false, but it
contains an element of truth: If you eschew prior pdfs, then you are
relegated to only ever asking and answering questions about the
probability of the data.  You can answer questions like ``what regions
of parameter space are consistent with the data?''\ but within the set
of consistent models, you can't answer questions like ``is this
parameter neighborhood more plausible than that one?''\note{It is hard
  to live as a frequentist but it is possible.  Some of the
  aforementioned battles between frequentists and Bayesians are set
  off by over-interpretation of a likelihood as a posterior pdf; in
  these cases the investigator is indeed unconsciously multiplying in
  a prior.  Other battles are set off by loose language in which a
  proper frequentist analysis is spoiled by over-statement of the
  results as making some parameters ``more probable'' than others.}
You also can't marginalize out nuisance parameters.

Although you might use the posterior pdf to report some kind of mean
prediction as discussed above, it almost never makes sense to just
\emph{optimize} the posterior pdf.  The posterior-optimal parameters
are called the ``maximum \notenglish{a posteriori}'' (MAP)
parameters.\note{The MAP parameters include prior information, which
  provide a measure on the parameters, so the MAP parameters change as
  the parameter space is transformed (as coordinate transformations
  are applied to the parameters).  This is not true of the maximum
  likelihood parameters, because the likelihood framework never makes
  use of any measure information.}  Like the maximum likelihood
parameters, these only make sense to compute if they are being used to
provide an approximation to the posterior pdf (in the form, say, of a
mode and a width).\note{I have seen many cases in which investigators
  report the MAP parameters instead of the maximum-likelihood (ML)
  parameters and describe the inference as ``Bayesian''.  That is
  pretty misleading, because neither the MAP nor the ML is a
  probabilistic output, and the MAP has additional assumptions in the
  form of a prior pdf.  That said, the MAP value is like a
  ``regularized'' version of the ML, and can therefore be valuable in
  engineering and decision systems; more about this elsewhere.}

The key idea is that the results of responsible data analysis is not
\emph{an answer} but a \emph{distribution over answers}.  Data are
inherently noisy and incomplete; they never answer your question
precisely.  So no single number---no maximum-likelihood or MAP
value---will adequately represent the result of a data analysis.
Results are always pdfs (or full likelihood functions); we must
embrace that.

Continuing our example of the model given in
\equationnames~(\ref{eq:example0}) through (\ref{eq:example1}), if we
want to learn the posterior pdf over the parameters $\theta \equiv [a,
  b]$, we need to choose a prior pdf $p(\theta \given I)$.  In
principle this should represent our actual prior or exterior
knowledge.  In practice, investigators often want to ``assume
nothing'' and put a very or infinitely broad prior on the parameters;
of course putting a broad prior is \emph{not} equivalent to assuming
nothing, it is just as severe an assumption as any other prior.  For
example, even if you go with a very broad prior on the parameter $a$,
that is a \emph{different} assumption than the same form of very broad
prior on $a^2$ or on $\arctan(a)$.  The prior doesn't just set the
ranges of parameters, it places a \emph{measure on parameter space.}
That's why it is so important.\note{The
  insertion of the prior pdf absolutely and necessarily increases the
  number of assumptions; there is no avoiding that.  Bayesians
  sometimes like to say that frequentists make just as many
  assumptions as Bayesians do.  It isn't true: A principled
  frequentist---an investigator who only uses the likelihood function
  and nothing else---genuinely makes fewer assumptions than any
  Bayesian.}

If you choose an infinitely broad prior pdf, it can become
\emph{improper}, in the sense that it can become impossible to satisfy
the normalization condition
\begin{eqnarray}\displaystyle
1 &=& \int p(\theta \given I)\,\dd\theta
\quad .
\end{eqnarray}
The crazy thing is that---although it is not advised---even with
improper priors you can still often do inference, because an
infinitely broad Gaussian (for example) is a well defined limit of a
wide but finite Gaussian, and the posterior pdf can be well behaved in
the limit.  That is, posterior pdfs can be proper even when prior pdfs
are not.

Sometimes your prior knowledge can be very odd.  For example, you
might be willing to take any slope $a$ over a wide range, but require
that the line $y=a\,x+b$ pass through a specific point
$(x,y)=(X_0,Y_0)$.  Then your prior might look like
\begin{eqnarray}\displaystyle
p(\theta \given I) &=& p(a \given I)\,p(b \given a, I)
\\\label{eq:dirac}
p(b \given a, I) &=& \delta(b + a\,X_0 - Y_0)
\quad ,
\end{eqnarray}
where $p(a \given I)$ is some broad function but $\delta(\cdot)$ is
the Dirac delta function, and, implicitly, $X_0$ and $Y_0$ are part of
the prior information $I$.  These examples all go to show that you
have an enormous range of options when you start to write prior pdfs.
In general, you should include all the things you know to be true when
you write your priors.  With great power comes great responsibility.

In other \documentnames\ in this series (for example
\citealt{straightline}), more advanced topics will be discussed.  One
example is the (surprisingly common) situation in which you have far
more parameters than data.  This sounds impossible, but the rules of
probability calculus don't prohibit it, and once you marginalize out
most of them you can be left with extremely strong constraints on the
parameters you care about.  Another example is that in many cases you
care about the \emph{prior} on your parameters, not the parameters
themselves.  Imagine, for example, that you don't know what prior pdf
to put on your parameters $\theta$ but you want to take it from a
family that itself has parameters $\beta$.  Then you can marginalize
out the $\theta$---yes, marginalize out all the parameters we once
thought we cared about---using the prior $p(\theta \given \beta, I)$
and be left with a likelihood for the parameters $\beta$ of the prior,
like so:
\begin{eqnarray}\displaystyle
p(D \given \beta, I) &=& \int p(D \given \theta, I)\,p(\theta \given \beta, I)\,\dd\theta
\quad .
\end{eqnarray}
The parameters $\beta$ of the prior pdf are usually called
``hyperparameters''; this kind of inference is called
``hierarchical''.\note{An example of simple hierarchical inference
  from my own work is \cite{eccentricity}.}

\begin{problem}\label{prob:gaussianprior}
Show that if you take the model in \equationnames~(\ref{eq:example0})
through (\ref{eq:example1}) and put a Gaussian prior pdf on $a$ and an
independent Gaussian prior pdf on $b$ that your posterior pdf for $a$
and $b$ will be a two-dimensional Gaussian.  Feel free to use informal
or even hand-waving arguments; there are no mathematicians present.
\end{problem}

\begin{problem}\label{prob:improper}
Take the limit of your answer to \problemname~\ref{prob:gaussianprior}
as the width of the $a$ and $b$ prior pdfs go to infinity and show
that you still get a well-defined Gaussian posterior pdf.  Feel free
to use informal or even hand-waving arguments.
\end{problem}

\begin{problem}
Show that the posterior pdf you get in \problemname~\ref{prob:improper}
is just a rescaling of the likelihood function by a scalar.  Two
questions: Why \emph{must} there be a re-scaling, from a dimensional
point of view?  What is the scaling, specifically?  You might have to
do some linear algebra, for which I won't apologize; it's awesome.
\end{problem}

\begin{problem}
\Equationname~(\ref{eq:dirac}) implies that the delta function
$\delta(q)$ has what dimensions?
\end{problem}

\clearpage
\markright{Notes}\theendnotes

\end{document}